\documentclass[12 pt]{article}\title{Mathematical Features of
continuity equation }\author{Saeed otarod\thanks{email:
sotarod\@yahoo.com}\\Department of Physics\\ Razi University
\\Kermanshah, Iran
}\begin{document}\maketitle\baselineskip=2\baselineskip\begin{abstract}
The conditions under which, the continuity equation can be
substituted by an ordinary non differential equation, will be
discussed. Since continuity equation is a fundamental equation,
this result will be applicable in a vast area of Physics.
\end{abstract} Subject headings: Mathematics
\cleardoublepage
\section{introduction}
In 2002 the author devised a method for solving nonlinear
differential equations[1]. Using this method one can solve many
important non linear differential equations in physics.\\ In an
endeavor to solve hydrodynamical equations governing the
interstellar media, we came to a very important result about
continuity equation. This result helped us to find exact solutions
for hydrodynamical equations in an straightforward and simple
way.\\In the following section we will write the result in the
form of a theorem. Although we have come to the result using the
new method, but in proving the theorem we do not use our method.
what we will do is simply substituting the driven result in
continuity equation.
\section{ Theorem}
For all density and velocity Functions in the Form, $\rho
=\rho(G(f(x,y,z,t)))$ and
$\overrightarrow{v}=\overrightarrow{v}(G(f(x,y,z,t)))$ the
continuity equation,  $\frac{\partial \rho}{\partial
t}+\nabla.(\rho \overrightarrow{v})=0$ can be written as an
ordinary equation if,  $f(x,y,z,t)=\alpha x+\beta y+\gamma z+\lambda t$.\\
 Here $G(f)$ is any arbitrary function of $f$.\\
 \textbf{Proof}:\\
 According to the stated conditions, Continuity equation can be written as,
 \begin{equation}
\frac{d \rho}{d G}\frac{d G}{d f}\frac{\partial f}{\partial
t}+\frac{dG}{d f}\frac{d(\rho \overrightarrow{v})}{d
G}.\overrightarrow{\nabla}f=0.
\end{equation}
As a result we will have,\begin{equation}\lambda \frac{d\rho}{d
G}+\frac{d}{d G}(\rho \overrightarrow{v}).(\alpha \hat{i}+\beta
\hat{j}+\gamma \hat{k})=0
\end{equation}
Integrating from both sides of the above equation will result in;
\begin{equation}
\rho+\rho \overrightarrow{v}.\hat{\kappa}=c
\end{equation}
c is a constant of integration and \begin{equation}
\hat{\kappa}=\alpha\hat{i}+\beta \hat{j}+\gamma
\hat{k}.\end{equation}
\\Therefore in these cases we can always write $\rho$ in terms of $\overrightarrow{v}$ as;
\begin{equation}
\rho=\frac{c}{1+\overrightarrow{v}.\hat{\kappa}}.
\end{equation}\\
\section{discussion}
Although very simple, we believe that this result is an important
result;
 Firstly, because we used this to find out explicit solution
for Naiver stokes equations in cases that had not been solved
previously(2).
 secondly, there is always a probability that, from explicit solutions
one may come to physical results that could not be driven out from
numerical analysis. At present, those who are concerned with
nonlinear physical applications, containing continuity equation,
use numerical methods for describing different physical phenomena.
One may check out the above result for new out comes.
\section{references}
 1-S.Otarod and
J.Ghanbari, Separation of variables for a nonlinear Differential
Equations. Electronic Journal of Differential Equations, Problem
section. 2002-2.\\
 2-S.Otarod and J.Ghanbari, Analytical Solution Of Hydrodynamical
 Equations. The common project of Razi  and Ferdowsi University.
 Phys.Depts of Razi and Ferdowsi University. Iran. 2003.
\end{document}